# Adaptive Frequency Response Reserve based on Real-time System Inertia

Shutang You

*Abstract*— To ensure adequate and economic reserve for primary frequency response in the current and future power system, this paper proposes real-time frequency response reserve (FRR) requirement based on system inertia. This minimum FRR will help power system operators adjust the current frequency response requirement and accommodate more renewable generations while achieving a saving of both energy and facility costs. Most importantly, the ability to adaptively vary the FRR will provide the additional agility, resiliency, and reliability to the grid.

## I. INTRODUCTION

The intermittence and fluctuation of renewable generations bring unprecedented challenges to power system reliability [1-17]. To keep lights on after resource contingencies such as generation trips, it is imperative for a power system to have sufficient frequency response reserve (FRR) to ensure that the frequency decay would not cross the frequency load shedding (UFLS) threshold [5, 18-35]. Besides real power, reactive power and voltage issues will arise due to the limited voltage regulation capabilities of inverters. Voltage levels will further influence the load consumption and inverter operation. Therefore, generation-load balance and frequency response will also be influenced by reactive power issues in renewable integration. However, since real power is more directly related to the system frequency, this paper mainly focuses on the active power's impact.

Frequency nadir is significantly influenced by system inertia, while maintaining an adequate FRR involves energy consumption and the consequent production cost increases [29]. Since the purpose of FRR is to keep the frequency higher than the UFLS threshold after a major contingency, the system inertia at the time of contingency should be taken into consideration in determining the minimum FRR [36].

In addition, some power systems' FRRs are maintained at a level much higher than necessary. As shown in Figure 1, the U.S. Eastern Interconnection (EI)'s frequency nadir is much higher than that of the ERCOT after a 1 GW generation loss contingency. Furthermore, the UFLS in the EI allows much smaller frequency excursions than most power grids in the world (as shown in Table 1). Australia and several other countries are considering further relaxing their UFLS thresholds for renewable integration since it will be merely uneconomical to maintain a very tight frequency requirement based on the worst scenario.

In practice, multiple strategies could be used to improve the frequency response [28]: for example, changing governor droop settings and dead-band settings of synchronous generation, battery energy storage with frequency droop control, super capacitor energy storage, etc. In current power grids, the frequency response capabilities of synchronous units still play an important role in providing primary frequency response to maintain frequency stability. Considering primary frequency response is an intrinsic capability of these synchronous units, it is more reasonable to leverage existing resources to the largest extent before using other alternative approaches. Therefore, this study focuses on FRR from synchronous generators.

Due to costs associated with FRR for synchronous generators, it will be beneficial for some power systems to dynamically adjust their FRR requirements. This is especially necessary as more renewables are connected to power grids. This paper uses the U.S. EI system as an example to first analyze the historical inertia values and determine the representative inertia cases. Then the minimum frequency response reserve in representative inertia conditions will be determined by dynamic simulation. Based on the minimum frequency response reserve requirement, this work will study how the system can derive real-time minimum frequency response reserve from inertia value. To further save frequency response reserve for the future high renewable system, the real-time minimum FRR requirement will be evaluated at a lower threshold of first-stage under-frequency-load-shedding (UFLS). Results on minimum real-time frequency response reserve will provide specific suggestions on future revision of the NERC BAL-003-1 standard.

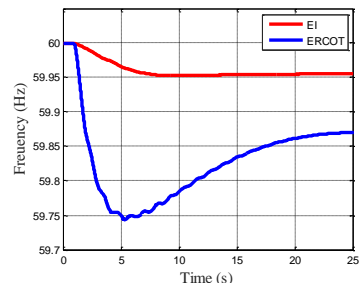

Figure 1. Frequency response of 1GW generation trip in EI and ERCOT

Table 1. First-stage UFLS of some power grids in the world

| Power Grid | Nominal Frequency (Hz) | First-stage UFLS threshold (Hz) |
|---|---|---|
| EI, US | 60 | 59.5 |
| ERCOT, US | 60 | 59.3 |
| Australian Energy Market Operator | 50 | 49 |
| National Grid, UK | 50 | 48.8 |
| Transpower, New Zealand | 50 | 47.8 |

## II. CORRELATION BETWEEN INERTIA AND FREQUENCY NADIR

After a contingency, the system frequency nadir is determined by the contingency magnitude, system inertia level, system governor response resources, and load damping effects, etc. As the system's largest contingency is usually pre-determined as the largest possible generation loss event, which is called Resource Contingency Criteria (RCC) by NERC. The initial decline rate of the system frequency is largely dependent on the system inertia, as presented in (1) [37].

$$\text{ROCOF} = f_N \cdot P_{Imbalance} / (2 H_{System} \cdot P_{System}) \quad (1)$$

where ROCOF is the rate of change of frequency. $P_{Imbalance}$ is the estimated power imbalance. $f_N$ is the nominal frequency (60 Hz). $P_{System}$ is the system total load. $H_{System}$ is the system inertia time constant. Figure 2 shows the change of system frequency nadir when the system inertia changes. It can be seen that lower system inertia will result in lower frequency nadir.

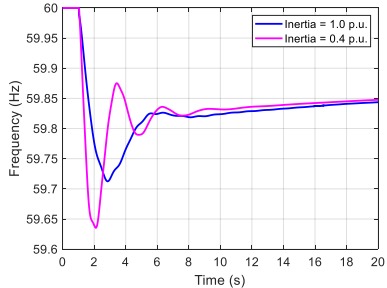

Figure 2. 18 system frequency response at different inertia levels

Inertia values will change with the system unit commitment, which is largely influenced by loads (as well as renewable output for high renewable systems). The system inertia will then influence system frequency stability. As an example, Figure 3 shows a projected profile of the change of system inertia in one day for an exemplar 18-bus system, whose maximum solar PV penetration is around 10%. It can be seen that the system inertia generally follows the change of the system load due to the low PV penetration rate. The system frequency nadir at each snapshot of this day is plotted in Figure 4. The two profiles show similar change patterns. However, their relationship is not linear due to the time constant and nonlinearity in frequency response. From these two graphs, it is easy to see that the system inertia has a significant impact on the frequency nadir.

To study the system frequency response reserve in different inertia levels, EI system historical and future projected inertia is studied as the first step. Figure 5 shows the EI system inertia curve at 15-minute time granularity and its probability distribution curve during 2016 to 2017. It can be seen that the diurnal variations are larger in peak load days than those in light load days. In addition, larger differences exist across different seasons compare with different hours within a day. The inertia distribution current EI system lies within the range of $1.2 \times 10^6$ to $2.4 \times 10^6$ MVA∗s.

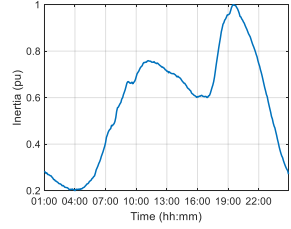 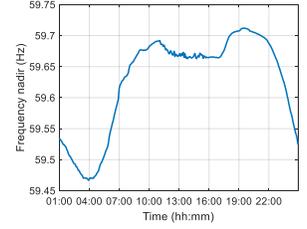

Figure 3. System inertia change in the 18-bus system in one day

Figure 4. System frequency nadir change of the 18-bus system in one day

EI Minimum Frequency Response Reserve under the Representative Inertia Conditions

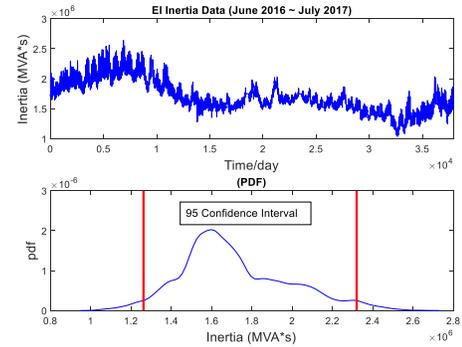

Figure 5. EI inertia curve and probability distribution function from June 2016 to July 2017 (15 minute time granularity)

Figure 6 and Figure 7 show the inertia distribution of the EI system when the system maximum PV penetration is 10% and 50%, respectively. It can be seen that the EI system inertia may fluctuate significantly on an hourly basis at high renewable penetration levels. As PV penetration increases, the inertia in the middle of a day will further decrease. The maximum and minimum inertia levels for 50% PV penetration are around $2.0 \times 10^6$ and $0.4 \times 10^6$ MVA ∗ s, respectively.

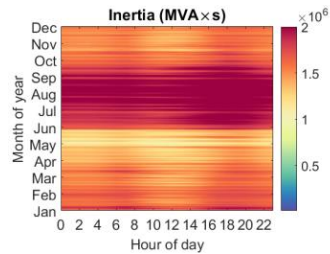 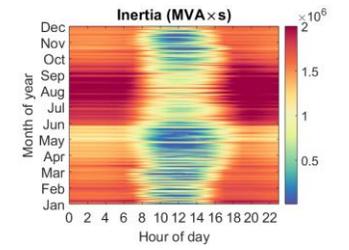

Figure 6. Inertia change in the EI system in one year with 10% PV penetration

Figure 7. Inertia change in the EI system in one year with 50% PV penetration

To study the change of frequency response reserve with the inertia level in the EI system, five models are developed for five representative conditions of the EI system. A summary of the five inertia levels is provided in Table 2.

Table 2. Studied inertia levels in the EI

| Scenario # | Renewable penetration | Inertia ( $1 \times 10^6$ MVA $*$ s) | % of base case inertia |
|---|---|---|---|
| 1 (base case) | 0% | 1.93 | N/A |
| 2 | 20% | 1.61 | 83.4% |
| 3 | 40% | 1.22 | 63.2% |
| 4 | 60% | 0.77 | 39.9% |
| 5 | 80% | 0.39 | 15.5% |

Figure 8 show the method to estimate FRR for each inertia level. Figure 9 shows the system frequency response at different FRR levels for Scenario 1. The FRR was gradually reduced until the frequency nadir crosses the UFLS threshold. The FRR was calculated by adding up headroom of all synchronous generators that can respond to frequency changes. Changing the limit of maximum real power output of synchronous generation reduce the headroom and FRR. In each step of FRR reduction, percentages of headroom reduction for each type of generators (hydro, gas, and coal, etc.) are kept the same.

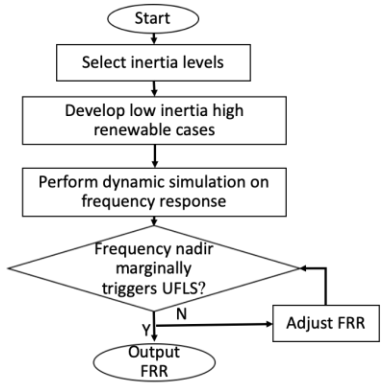

Figure 8. The method to calculate FRR for various inertia levels.

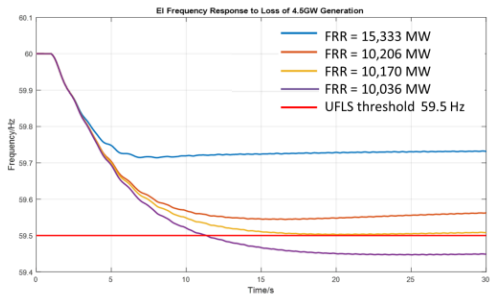

Figure 9. Adjust FRR until reaching the UFLS threshold (Scenario 1)

Figure 10 shows the change of minimum frequency-responsive generation capacity with system inertia in the EI. This result shows that FRR increases between 2,000-3,000 MW for every $1 \times 10^5$ MVA*s reduction in system inertia. In addition, the relation between the minimum FRR and the EI inertia is non-linear: the minimum FRR increases faster as the system inertia reduces. This information helps EI operators to procure and deploy FRR resources to ensure system reliability while minimizing costs.

To estimate the potential savings of adjusting frequency response reserve based on real-time inertia, a day scenario was created in which system inertia can reduce to 20% of its initial value at the maximum instantaneous renewable penetration (80%) period. Figure 11 shows the PV output during one cloudy day in the EI. Figure 12 shows the EI inertia changes with PV output during this day. The Business-As-Usual (BAU) curve, which represents a low solar PV penetration case, was also plotted for comparison. It can be seen that the high PV output in the middle of the day will result in low system inertia.

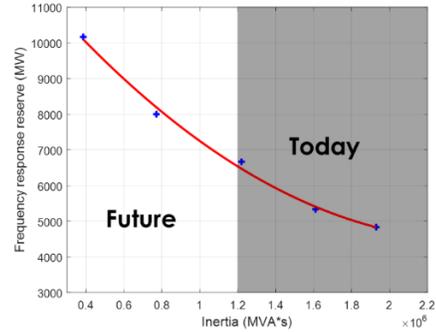

Figure 10. Change of FRR at different inertia levels

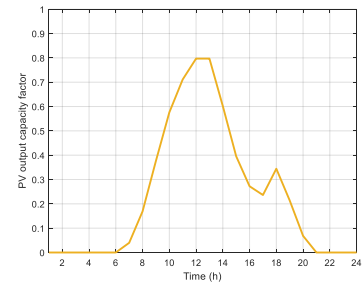

Figure 11. PV output capacity factor during one day

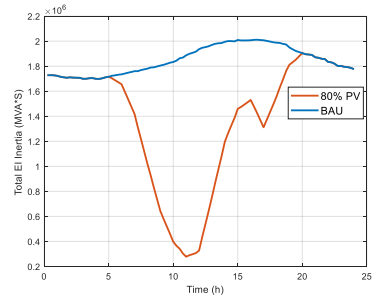

Figure 12. Inertia change of the EI with 80% maximum PV penetration during one day

Based on the system inertia curve in Figure 12 and the relation between inertia and the FRR provided in Figure 10, the real-time system FRR can be obtained. Figure 13 shows the change of hourly frequency reserve power requirement in this day. It can be seen that the peak of frequency response reserve also happens around the middle of the day. The reduction in the FRR indicates potential savings in the resources. Although primary frequency response is not a standalone ancillary market product due to low penetration of renewables in the current EI system, the primary frequency response market will probably be necessary for the future high renewable EI system and is under consideration in many other power grids. Since primary frequency response and the regulation spinning reserve require the same headroom, the cost saving is assessed using the average real-time market clearance price of regulation spinning reserve for three representative ISOs in the EI.

Table 3. ISO real-time market clearance price of regulation reserve

| ISO | Price ($/MW) |
|---|---|
| NYISO [38] | 12.00 |
| PJM [39] | 15.92 |
| MISO [40] | 12.76 |
| Average | 13.56 |

Figure 14 shows the cost savings with varying FRR. In this study scenario of the EI, using real-time inertia to adjust frequency response reserve can potentially save more than 40% cost in future primary frequency response energy.

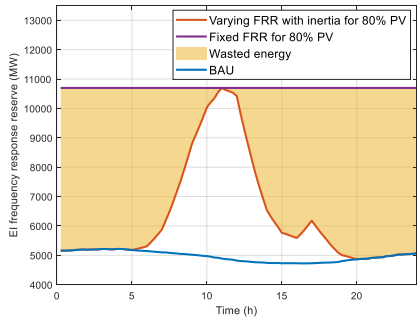

Figure 13. Frequency response reserve values in the EI with 80% maximum PV penetration

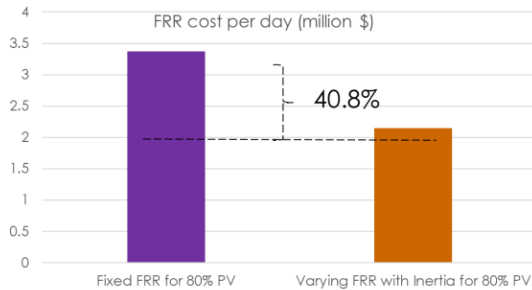

Figure 14. Frequency response reserve cost savings

Similarly, for a lower UFLS threshold, frequency FRR values can be calculated using the same approach. Figure 15 shows a comparison of frequency response reserve values using 59.3 Hz versus 59.5 Hz as the first-stage UFLS threshold. It can be seen that lower UFLS threshold will relax the requirement on frequency response reserve. The difference in reserve is not very large. The reason is that although less frequency-responsive generators are needed for a lower UFLS threshold, but each frequency-responsive generator needs to have larger headroom due to larger frequency deviation.

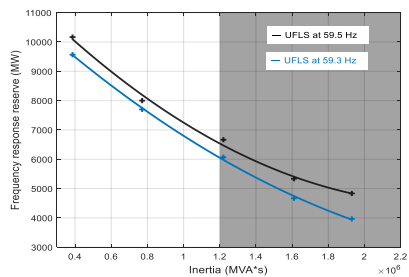

Figure 15. Comparison of frequency response reserve values using 59.3 Hz versus 59.5 Hz as the UFLS threshold

## III. RECOMMENDATIONS TO FUTURE REVISION OF STANDARDS

NERC Standard BAL-003-1 [41, 42] contains Interconnection Frequency Response Obligation (IFRO) for each interconnection grid in the U.S. The increase of renewable generation will change the requirement on how much frequency response reserve the system should have to ensure system frequency stability. Besides real-time FRR based on system inertia, results of this study also suggest some additional changes on future standard to accommodate increasing renewables and system inertia reduction.

### A. Frequency response obligation directly calculated at Point C.

Figure 16 show the system frequency response for the three main interconnection grids in the U.S measured by GridEye [43-67]. It can be seen that point C is much smaller than point B for low inertia grids (ERCOT and WECC). For these low inertia power grids, Point C is more critical than Point B. PMU measurements at generators and tie-lines may be needed to improve accountability of frequency response services from generators and BAs.

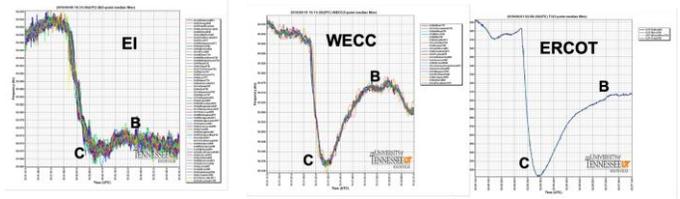

Figure 16. Frequency responses in three U.S. interconnections from GridEye [17, 49, 52, 58, 64, 68]

### B. Resource Contingency Criteria (RCC) magnitude may need to be updated with the increase of renewable penetration.

The unit commitment and dispatch pattern will change with the increase of renewables. Conventional generation units tripped in the RCC event may have been retired so that the event magnitude may need to be updated. In addition, inverter-based renewable generation may induce other grid events due to their technical features, such as protection schemes, ramping and fluctuations, voltage and frequency support characteristics, etc. For example, the Blue Cut fire caused faults in transmission lines and consequently resulted in 1200 MW PV generation loss due to voltage drops and the disturbed frequency signal [69]. This is a new type of events related to renewable generation that may need to be considered.

### C. Redesign Balancing Authority (BA) frequency response requirement.

The frequency response capability of each BA largely depends on the generation mix and the frequency response characteristics of each type of generation. Since the distribution of renewable generation resources is uneven, the distribution of frequency response capabilities across BAs become more unbalanced as more renewable generation is integrated to the system. It becomes difficult for BAs with high renewable penetration to meet the frequency response requirement compared with BAs with low renewables. In the future, BA frequency response may be allowed to trade in the FRR market to meet BA-level obligations. Figure 17 shows the change of regional frequency response in one future high PV penetration scenario of the EI. The regional frequency response capabilities decrease as the PV penetration increases. Regions with high PV penetration, such as PJM, NYISO, and NEISO, will not fulfill the frequency response obligations as the renewable penetration reaches 80%. In contrast, BAs with low PV penetration, for example, TVA and SOCO, will satisfy the frequency response requirements even at 80% renewable penetration level because of frequency-responsive hydro power plants in these regions.

In high renewable penetration grids, the diversity in frequency response capabilities indicates a need to distinguish each BA in frequency response obligations (FROs) based on its renewable penetration, as well as an opportunity to trade FRO across BAs with different penetration rates of renewables.

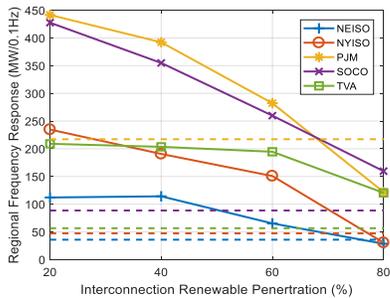

Figure 17. Change of regional frequency response in the EI (Dash lines represent $FRO_{BA}$ calculated using the method suggested by BAL-003)

IV. CONCLUSIONS

This paper analyzed the current inertia data and projected future inertia levels of the EI system. Minimum frequency response reserves were calculated using dynamic simulations at multiple inertia levels as representative conditions in the current and future high renewable EI system. A one-day scenario with 80% peak renewable penetration was developed to test the performance of proposed real-time FRR in saving energy costs and ensuring reliability. The study result shows that using real-time inertia to adjust FRR can potentially save more than 40% cost in future primary frequency response energy. Minimum FRR values when first-stage UFLS threshold reduces to 59.3 Hz were obtained and compared with minimum FRR values at the default 59.5 Hz. This work also provided suggestions on potential changes on NERC BAL-003-1 to accommodate future high renewable in the EI system.